# Decision Support to Crowdsourcing for Annotation and Transcription of Ancient Documents: The RECITAL Workshop

*Olivier Aubert (contact@olivieraubert.net), LS2N ; Université de Nantes, France*

*Benjamin Hervy (benjamin.hervy@univ-nantes.fr), Centre François Viète ; Université de Nantes, France*

*Guillaume Raschia (guillaume.raschia@univ-nantes.fr), LS2N ; Université de Nantes, France*

*Françoise Rubellin (francoise.rubellin@univ-nantes.fr), LAMO ; Université de Nantes, France*

## Introduction

In the 18th century in Paris, only two public theatres could officially perform comedies: the Comédie-Française, and the Comédie-Italienne. The latter was much less well known. By studying a century of accounting registers, we aim to learn more about its successful plays, its actors, musicians, set designers, and all the small trades necessary for its operation, its administration, logistics and finances. We believe that the material and logistical history of the Comédie-Italienne holds great surprises in comparison with academic prejudices (Rubellin / Raschia 2020).

To this end, we employ a mass of untapped and unpublished resources, the 27,544 pages of 63 daily registers available at the Bibliothèque Nationale de France (BnF). And we take a decidedly fresh look at emerging forms of creation and changes in the entertainment economy. In parallel to Handwritten Text Recognition technology (Granet et al. 2018), we developed the crowdsourcing platform RECITAL to collect and index the data from the registers, following an emerging trend in Digital Humanities (Terras 2016).

RECITAL is built upon the scribeAPI framework and it offers a fully-fledged web application to classify the pages, annotate with marks and tags, transcribe the indexed marks and even to verify the previous transcripts. All those features are open to volunteers visiting the web site.

One of the main challenges that comes after the crowdsourcing process is to end up with a complete and reliable database, despite the non-controlled work from volunteers and the required post-processing (Causer et al. 2018) (Blickhan et al. 2019). Our contribution to that open issue was to design a multi-level data model and to develop a series of monitoring and decision tools to support crowdsourced data management up to their definitive form.

## The Four-Staged Data Model

The very first data model, so-called CrowdSourcing (or CS) Model, is driven by the general-purpose crowdsourcing system, powered by the scribeAPI framework. It mainly records crowdsourced data as a large (more than 300k entries) log of volunteer task runs, with a generic model featuring Subjects (micro-tasks) and Classifications (volunteer actions).

Then, an Extract-Transform-Load (ETL) process turns the CS Model into a Register-Page-Mark-Transcript chain of entities, coined the Raw Model. This data exchange process is a one-to-one mapping and does not distort, aggregate or enrich the data. It changes the point of view from a large log of micro-task runs coming from the volunteer's work to a « physical », i.e., artifact-oriented, representation. As a side-effect, the exchange process, from the CS model to the Raw model, helps to fix misbehavior of the CS platform with respect to the complex task assignment policy. Both CS and Raw models are populated with an append-only strategy.

Next, raw data are cleansed thanks to an intensive automated post-processing step (Hervy et al. 2019) including natural language processing (NLP), record linkage, and inter-annotator agreement to reach consensus. The expected result of that second step is (i) a subset of page categories, marks, indexes and transcriptions tagged as fully confident, (ii) another subset being almost confident and (iii) everything else (highly questionable or outlier data). The output is recorded into the so-called Cooked Model.

A PROV-like mechanism is set up to keep track of provenance for every piece of data inserted into the Cooked Model and to provide information aimed at estimating its reliability. Obviously, the raw data are neither deleted, nor updated during this step. Dependencies are created only from an upper layer to its proper underlying model. Modeling provenance and confidence as additional metadata has significant value considering the epistemological aspect of this work.

The fourth stage of the data model is made of entities and relationships related to the historical material, that is to say, the day after day Comédie-Italienne shows all along the 18th century and the commedia dell'arte. This is the ultimate Domain Model that is exposed to historians, literarians, musicologists and economists to name a few.

The four-staged model is the necessary foundation to address scientific, technical and even epistemological challenges among which: monitoring progress, assessing data quality, cleaning and validating data, and accepting data. However, it still requires to be equipped with access facilities and practical user interfaces in order to be fully operational.

## Monitoring and Decision Tools

To support the overall transformation process from the CS Model to the Domain Model, we proposed a series of analytics and interactive tools called the RECITAL workshop. It is primarily addressed to « humanists » in order for them to explore intermediate live states of the data collection process and provide them with a way to drive the remaining work.

Crowdsourcing systems usually give key performance indicators such like task completeness and volunteer's activity to monitor an overall project. Coordination and communication between the requester and the volunteers remain open issues (Bhatti et al. 2020). Furthermore, ad hoc post-processing techniques, including human validation, are applied with few to none documentation and even less reproducibility options. The RECITAL workshop brings transparency to that all « behind the scene » work. It first exposes the entire data model through a REST web server. Then the dashboard, powered by a web application, presents the data with the many index entries such as register, page and marks, transcript, volunteer, play, show, actor, etc, displaying their information alongside the original image data. It can then be used by computer scientists as well as humanities researchers to explore and assess the raw data (directly obtained through the crowdsourcing process) as well as the cooked data (post-processed, cleansed data). Providing access to the various layers of information, it can be used to follow data transformations – using the PROV-like information – in order to under-stand possible inaccuracies in the processing pipeline.

In addition to the raw display of the transcribed information on the original pages, the dashboard produces digital surrogates of the original documents, either graphical reconstitutions of the manuscripts with a similar layout, or raw text-based reconstitutions of the manuscripts, more oriented towards the actual content rather than its layout.

All those features aim at supporting researchers towards a better understanding and acceptance of the data transformation pipeline, and also, to grasp the historical content itself, providing deep insights into the documents in a live and interactive manner.

## References

Enter your references here:

Special Issue on Collecting, Preserving, and Disseminating Endangered Cultural Heritage for New Understandings through Multilingual Approaches, December.

**Causer, Tim** / **Grint, Kris** / **Sichani, Anna-Maria** / **Terras, Melissa** (2018): « Making such Bargain: Transcribe Bentham and the Quality and Cost-Effectiveness of Crowdsourced Transcription », in: *Digital Scholarship in the Humanities* 33, 3: 467–487.

**Granet, Adeline** / **Hervy, Benjamin** / **Roman Jimenez, Geoffrey** / **Hachicha, Marouane** / **Morin, Emmanuel** / **Mouchère, Harold** / **Quiniou, Solen** / **Raschia, Guillaume** / **Rubellin, Françoise** / **Viard-Gaudin, Christian** (2018): « Crowdsourcing-based Annotation of the Accounting Registers of the Italian Comedy », in: *11th International Conference on Language Resources and Evaluation (LREC)*. Miyazaki, Japan, May.

**Hervy, Benjamin** / **Pétillon, Pierre.** / **Pigeon, Hugo** / **Raschia, Guillaume** (2019): « Data Correction for Transcription in Crowdsourcing. A Feedback from the RECITAL Platform », in: *Information Retrieval, Document and Semantic Web* 2, 1. ISSN 2516-3280. doi: 10.21494/ISTE.OP.2019.0348.

**Rubellin, Françoise** / **Raschia, Guillaume** (2020): « Redécouvrir les théâtres de la Foire et la Comédie-Italienne avec les bases THEAVILLE et RECITAL », in: *Revue d'Historiographie du Théâtre* 5.

**Terras, Melissa** (2016): « Crowdsourcing in the Digital Humanities », in: Schreibman, Susan / Siemens, Ray / Unsworth, John (eds): *A New Companion to Digital Humanities*. John Wiley & Sons, Ltd: 420-438, Oxford, UK. ISBN 9781118680605.